\documentclass[trackchanges,twocolumn]{aastex7}
\usepackage{amsmath}
\usepackage{soul} 
\usepackage{bm}
\usepackage{booktabs}
\usepackage{subcaption} 
\usepackage{amsmath}
\usepackage{units}

\begin{document}

\title{Searching for continuous gravitational waves in the Parkes Pulsar Timing Array DR3}

\author[0009-0001-8885-5059]{Shi-Yi Zhao}\email{zhaosyastro@gmail.com}
\affiliation{School of Physics and Astronomy, Beijing Normal University, Beijing 100875, China}
\affiliation{Department of Physics, Faculty of Arts and Sciences, Beijing Normal University, Zhuhai 519087, China}

\author[0000-0001-7016-9934]{Zu-Cheng Chen}
\email{zuchengchen@hunnu.edu.cn}
\affiliation{Department of Physics and Synergetic Innovation Center for Quantum Effects and Applications, Hunan Normal University, Changsha, Hunan 410081, China}
\affiliation{Institute of Interdisciplinary Studies, Hunan Normal University, Changsha, Hunan 410081, China}

\author[0000-0001-9852-6825]{Jacob Cardinal Tremblay}
\email{jacobcardinaltremblay@gmail.com}
\affiliation{Max Planck Institute for Gravitational Physics (Albert Einstein Institute), 30167 Hannover, Germany}
\affiliation{Leibniz Universität Hannover, 30167 Hannover, Germany}

\author[0000-0003-3189-5807]{Boris Goncharov}
\email{boris.goncharov@gssi.it}
\affiliation{Max Planck Institute for Gravitational Physics (Albert Einstein Institute), 30167 Hannover, Germany}
\affiliation{Leibniz Universität Hannover, 30167 Hannover, Germany}

\author[0000-0001-7049-6468]{Xing-Jiang Zhu}
\affiliation{Department of Physics, Faculty of Arts and Sciences, Beijing Normal University, Zhuhai 519087, China}
\affiliation{Institute for Frontier in Astronomy and Astrophysics, Beijing Normal University, Beijing 102206, China}
\email[show]{zhuxj@bnu.edu.cn}

\author[0000-0002-8383-5059]{N. D. Ramesh Bhat}
\email{rameshbhatnd@gmail.com}
\affiliation{International Centre for Radio Astronomy Research, Curtin University, Bentley, WA 6102, Australia}

\author[0000-0002-7031-4828]{Ma\l{}gorzata Cury\l{}o}
\email{mcurylo@astrouw.edu.pl}
\affiliation{School of Physics and Astronomy, Monash University, Clayton VIC 3800, Australia}
\affiliation{Australian Research Council Centre of Excellence for Gravitational Wave Discovery (OzGrav)}

\author[0000-0002-9618-2499]{Shi Dai}
\email{Shi.Dai@csiro.au}
\affiliation{Australia Telescope National Facility, CSIRO, Space \& Astronomy, PO Box 76, Epping, 1710, NSW, Australia}
\affiliation{Western Sydney University, Locked Bag 1797, Penrith South DC, NSW 2751, Australia}

\author[0000-0003-3432-0494]{Valentina Di Marco}
\email{valentina.dimarco@monash.edu}
\affiliation{School of Physics and Astronomy, Monash University, Clayton VIC 3800, Australia}
\affiliation{Australian Research Council Centre of Excellence for Gravitational Wave Discovery (OzGrav)}
\affiliation{Australia Telescope National Facility, CSIRO, Space \& Astronomy, PO Box 76, Epping, 1710, NSW, Australia}

\author[0000-0002-9174-638X]{Hao Ding}
\email{hdingastro@hotmail.com}
\affiliation{Mizusawa VLBI Observatory, National Astronomical Observatory of Japan, 2-12 Hoshigaoka-cho, Mizusawa, Oshu, Iwate 023-0861, Japan}

\author[0000-0003-1502-100X]{George Hobbs}
\email{George.Hobbs@csiro.au}
\affiliation{Australia Telescope National Facility, CSIRO, Space \& Astronomy, PO Box 76, Epping, 1710, NSW, Australia}

\author[0009-0001-5071-0962]{Agastya Kapur}
\email{agastya.kapur@students.mq.edu.au}
\affiliation{Australia Telescope National Facility, CSIRO, Space \& Astronomy, PO Box 76, Epping, 1710, NSW, Australia}
\affiliation{Department of Mathematics and Physical Sciences, Macquarie University, NSW 2109, Australia}

\author[0009-0009-9142-6608]{Wenhua Ling}
\email{wenhua.ling@csiro.au}
\affiliation{Australia Telescope National Facility, CSIRO, Space \& Astronomy, PO Box 76, Epping, 1710, NSW, Australia}

\author[0000-0002-5248-5076]{Tao Liu}
\email{taoliu@ust.hk}
\affiliation{Department of Physics and Jockey Club Institute for Advanced Study, The Hong Kong University of Science and Technology, Hong Kong S.A.R., China}

\author[0000-0001-5131-522X]{Rami Mandow}
\email{rami.mandow@hdr.mq.edu.au}
\affiliation{Department of Mathematics and Physical Sciences, Macquarie University, NSW 2109, Australia}
\affiliation{Australia Telescope National Facility, CSIRO, Space \& Astronomy, PO Box 76, Epping, 1710, NSW, Australia}

\author[0009-0001-5633-3512]{Saurav Mishra}
\email{sauravmishra1206@gmail.com}
\affiliation{Centre for Astrophysics and Supercomputing, Swinburne University of Technology, P.O. Box 218, Hawthorn, VIC, 3122, Australia}
\affiliation{Australian Research Council Centre of Excellence for Gravitational Wave Discovery (OzGrav)}
\affiliation{Australia Telescope National Facility, CSIRO, Space \& Astronomy, PO Box 76, Epping, 1710, NSW, Australia}

\author[0000-0002-2035-4688]{Daniel J. Reardon}
\email{dreardon@swin.edu.au}
\affiliation{Centre for Astrophysics and Supercomputing, Swinburne University of Technology, P.O. Box 218, Hawthorn, VIC, 3122, Australia}
\affiliation{Australian Research Council Centre of Excellence for Gravitational Wave Discovery (OzGrav)}

\author[0000-0002-1942-7296]{Christopher J Russell}
\email{chris.russell@csiro.au}
\affiliation{CSIRO Scientific Computing, Australian Technology Park, Locked Bag 9013, Alexandria, NSW 1435, Australia}

\author[0000-0002-7285-6348]{Ryan M. Shannon}
\email{rshannon@swin.edu.au}
\affiliation{Centre for Astrophysics and Supercomputing, Swinburne University of Technology, P.O. Box 218, Hawthorn, VIC, 3122, Australia}
\affiliation{Australian Research Council Centre of Excellence for Gravitational Wave Discovery (OzGrav)}

\author[0000-0003-4498-6070]{Shuangqiang Wang}
\email{wangshuangqiang@xao.ac.cn}
\affiliation{Xinjiang Astronomical Observatory, Chinese Academy of Sciences, Urumqi, Xinjiang 830011, China}
\affiliation{Australia Telescope National Facility, CSIRO, Space \& Astronomy, PO Box 76, Epping, 1710, NSW, Australia}

\author[0000-0001-8539-4237]{Lei Zhang}
\email{leizhang996@nao.cas.cn}
\affiliation{National Astronomical Observatories, Chinese Academy of Sciences, Beijing 100101, China}
\affiliation{Centre for Astrophysics and Supercomputing, Swinburne University of Technology, P.O. Box 218, Hawthorn, VIC, 3122, Australia}

\author[0000-0002-9583-2947]{Andrew Zic}
\email{andrew.zic@csiro.au}
\affiliation{Australia Telescope National Facility, CSIRO, Space \& Astronomy, PO Box 76, Epping, 1710, NSW, Australia}
\affiliation{Australian Research Council Centre of Excellence for Gravitational Wave Discovery (OzGrav)}

\begin{abstract}

We present results from an all-sky search for continuous gravitational waves from individual supermassive binary black holes using the third data release (DR3) of the Parkes Pulsar Timing Array (PPTA). Even though we recover a common-spectrum stochastic process, potentially induced by a nanohertz gravitational wave background, we find no evidence of continuous waves. Therefore, we place upper limits on the gravitational-wave strain amplitude: in the most sensitive frequency range around 10 nHz, we obtain a sky-averaged 95\% credibility upper limit of $\approx 7 \times 10^{-15}$. Our search is sensitive to supermassive binary black holes with a chirp mass of $\geq 10^9M_{\odot}$ up to a luminosity distance of 50 Mpc for our least sensitive sky direction and 200 Mpc for the most sensitive direction. This work provides at least 4 times better sensitivity in the 1-200 nHz frequency band than our last search based on the PPTA's first data release.
We expect that PPTA will continue to play a key role in detecting continuous gravitational waves in the exciting era of nanohertz gravitational wave astronomy.

\end{abstract}



\section{Introduction}
\label{sec:intro}

The primary goal of pulsar timing arrays (PTAs) is to detect and characterize nanohertz frequency gravitational waves.
After several decades of efforts, there has been significant recent progress in detecting a stochastic gravitational wave background (GWB).
It began with the identification of an excess noise that is common among the times of arrival measurements for different pulsars \citep{NG_12_CRN,PPTA_DR2_CRN,EPTA_CRN,IPTA_DR2_CRN,crn_pol, crn_romano}.
This was referred to as common red noise or a common-spectrum process and was thought to be caused by a GWB, for which the direct evidence would be the Hellings-Downs \citep[HD;][]{hellings1983upper} spatial correlation.
What followed was indeed the discovery of the HD correlation (with relatively low statistical significance) found in the data sets of the North American Nanohertz Observatory for Gravitational Waves \citep[NANOGrav;][]{agazie2023nanograv}, the PPTA \citep{reardon2023search}, the European PTA (EPTA) in combination with the Indian PTA (InPTA) \citep{antoniadis2023second}, the Chinese PTA \cite[CPTA;][]{xu2023searching}, and more recently the MeerKAT PTA \citep{MPTA25gwb}.

The leading source of a GWB in the nHz frequency band is the combined emission from supermassive binary black holes (SMBBHs) distributed across the Universe.
Within such a cosmic population of SMBBHs, the nearest and/or the most massive binaries can be detected individually in the form of continuous gravitational waves (CGWs); see also \citet{HIGH_RS} for the possibility of detecting high-redshift binaries.
Over the next decade, once the detection of a GWB is confirmed, the following biggest question is: what is the origin of this GWB?
The discovery of a single SMBBH would not only demonstrate that the background is likely made from multiple SMBBH sources, but also enable the properties of the binary system to be measured through possibly multi-messenger observations \citep[see, e.g.,][]{LiuTT21,XinCC21,PetrovTaylor24,TruantSesana25}.

Using the 64-m Parkes radio telescope, the PPTA started regular timing observations of 20 millisecond pulsars from early 2004.
Three data sets have been published, with a steadily growing sample of pulsars and increasingly higher timing precision \citep{pptadr1,pptadr2,zic2023parkes}.
Our last search for CGWs was performed on the PPTA DR1 \citep{zhu2014all}.
With improved sensitivities, more recent searches with other PTA data sets produced interesting limits on potential SMBBHs in nearby galaxies \citep{NG21nearby}, constraints on candidate SMBBHs found in electromagnetic-wave surveys \citep{NG243c66b}, in addition to tighter upper limits on the strain amplitude of CGWs \citep{eptadr2cw, agazie2023nanogravcw}.

In this paper, we use the PPTA DR3 \citep{zic2023parkes} to conduct an all-sky blind search for CGWs generated by single SMBBHs. We consider only binary systems with circular orbits\footnote{
While there is a motivation to search for CGWs from binaries in eccentric orbits~\citep{taylor2016ecc}, our search is also effective for the initial detection of eccentric binaries. 
See, \textit{e.g.}, Figure~11 in~~\cite{taylor2016ecc} and the conclusions in~\cite{ZhuWen2015}.}.
For each pulsar, we employ the noise model published in \citet{reardon2023search}.
The plan of this paper is as follows. In Section \ref{sec:methods}, we briefly describe the data set, introduce the signal and noise models, and give an overview of the analysis methods used in this study. Section \ref{sec:result} includes the search results and derived upper limits on individual SMBBHs.
Finally, we present conclusions in Section \ref{sec:conclusions}.

\section{Data and Methods} \label{sec:methods}

\subsection{The PPTA DR3} \label{subsec:data}

The PPTA DR3 data set includes high-precision timing observations of 32 millisecond pulsars, obtained using the Parkes radio telescope from February 6, 2004, to March 8, 2022, with a typical cadence of three weeks. Details of this data set and a comprehensive noise analysis are presented in \citet{zic2023parkes} and \citet{reardon2023noises}, respectively. The data set is publicly available at the CSIRO Data Access Portal \citep{PPTA_DR3_data}. The PPTA DR1 includes observations of 20 pulsars that span 6 years, which were extended to $\sim 14$ years and 26 pulsars in DR2. The DR3 data set contains an updated version of DR2 and over 3 years of observations obtained using the ultrawide-bandwidth low-frequency receiver \citep[UWL;][]{parkes_uwl}. 

Following \citet{reardon2023search}, we only use data from 30 pulsars in this work. The globular cluster pulsar J1824$-$2452A was excluded as it contains steep-spectrum red noise that is too strong for this pulsar to be sensitive to a common process. The noise is likely to be intrinsic to the pulsar \citep{shannon2010assessing}, although globular-cluster dynamics may play a role as well. We also exclude PSR J1741+1351 from the analysis. It was only added to the PPTA as the UWL was commissioned and observed with low priority, leading to a small number of observations.

In this work, we make use of noise models developed in \citet{reardon2023noises}. For our analysis, we use the DE440 solar system ephemeris (SSE) \citep{DE440} and adopt the Terrestrial Time (TT)(BIPM2020) reference timescale. In the following subsections, we present our noise and signal models.

\subsection{Noise models} \label{subsec:model}

In our model, pulsar timing residuals consist of the following components: 
\begin{equation}
\delta t=M\epsilon+n_\text{PSRN}+n_\text{CRN}+s,
\end{equation}
where $M$ is the design matrix of the linearized timing model, $\epsilon$ represents the timing model parameter offsets, $n_\text{PSRN}$ and $n_\text{CRN}$ are the pulsar-specific noise \citep{reardon2023noises} and a common red noise (CRN), respectively, and $s$ represents the CGW signals.

A CRN refers to a noise process that shares a common red power spectrum across all pulsars; it was first reported in GWB searches with NANOGrav 12.5-year data set\citep{NG_12_CRN}, the PPTA DR2 \citep{PPTA_DR2_CRN}, EPTA \citep{EPTA_CRN}, and IPTA DR2 \citep{IPTA_DR2_CRN}. Stronger evidence was found for a CRN in latest PTA data sets.
We include a CRN term in our noise model for the CGW analysis.
The power spectral density of the CRN is given by:
\begin{equation}
P_\text{CRN}(f;A_\text{CRN},\gamma_\text{CRN})=\frac{A_\text{CRN}^2}{12\pi^2} f_\mathrm{yr}^{-3} \left( \frac{f}{f_{\mathrm{yr}}} \right)^{-\gamma_\text{CRN}},
\end{equation}
where $A_\text{CRN}$ is the amplitude, and $\gamma_\text{CRN}$ is the power-law spectral index. The prior on $A_\text{CRN}$ is log-uniform in the range $[-18,-11]$, while the prior on $\gamma_\text{CRN}$ is uniform in the range $[0, 7]$. Similar to \citet{reardon2023search}, the CRN is estimated in $[T_{\rm obs}/(240 \text{ days})]=28$ linearly spaced frequency bins, ranging from $1/T_{\rm obs}$ to $28/T_{\rm obs}$, where $T_{\rm obs}=6605$ days is the total observing span of PPTA DR3. This division is because the sensitivity of PPTA DR3 is dominated by white noise near $f \thicksim 1/(240 \text{ days})$ as determined by a broken power-law analysis \citep{reardon2023search}.

For pulsar-specific noise $n_\text{PSRN}$, the first component is white noise caused by observational uncertainties. We use a white noise model consistent with previous PPTA analyses, characterized by three parameters: EFAC, EQUAD, and ECORR. These parameters are applied independently to each unique pulsar receiver-backend system. EFAC scales the template-fitting time of arrival (TOA) uncertainties induced by finite pulse signal-to-noise ratios by a multiplicative factor. EQUAD adds white noise in quadrature, and ECORR describes white noise that is correlated across TOAs derived from data collected simultaneously.

For $n_\text{PSRN}$, we consider a number of spectrum-dependent noise components, including achromatic red noise (Red), dispersion measure (DM) variations, high fluctuation frequency (HFF) noise, chromatic noise (Chr), and band noise (BN). We adopt a power-law model for the power spectral density of these noise terms \citep{shannon2010assessing, lam2016nanograv, reardon2023noises}:

\begin{equation}
P(f;A,\gamma)=\frac{A^2}{12\pi^2}f_\text{yr}^{-3}\left(\frac{f}{f_{\text{yr}}}\right)^{-\gamma}.
\end{equation}
The specific spectral indices ($\gamma$) and amplitudes ($A$) for these noise components can be found in Table 1 of \citet{reardon2023noises}.
Additional details of noise analyses for our data set are presented in \citet{reardon2023noises}.

\subsection{Continuous gravitational waves} \label{subsec:CGW}
The timing residuals induced by CGW can be written as \citep[see, e.g.,][]{zhu2014all}: 
\begin{equation}
\label{eq:residual}
s(t)=F^+(\hat{\Omega})[s_{+}(t_{p})-s_{+}(t)]+F^{\times}(\hat{\Omega})[s_\times(t_p)-s_\times(t)],
\end{equation}
here $F(\hat{\Omega})^{+,{\times}}$ are the antenna pattern functions for the plus and cross polarizations; they describe the pulsar's response to a CGW coming from a direction $\hat{\Omega}$:

\begin{align}
F^+(\hat{\Omega}) &= \frac{1}{4(1-\cos{\alpha})} \big\{ 
(1+\sin^2 \phi)\cos^2 \phi_p \cos[2(\theta_p-\theta)] \nonumber \\
&\quad - \sin 2\phi \sin 2\phi_p \cos(\theta_p-\theta) \nonumber \\
&\quad + \cos^2 \phi (2 - 3\cos^2 \phi_p) 
\big\}
\end{align}

\begin{align}
F^{\times}(\hat{\Omega}) &= \frac{1}{2(1-\cos{\alpha})} \big\{ 
\cos \phi \sin 2\phi_p \sin(\theta_p - \theta) \nonumber \\
&\quad - \sin \phi \cos^2 \phi_p \sin[2(\theta_p - \theta)] 
\big\}
\end{align}
where $(\theta,\phi)$ denotes the sky coordinates of the SMBBH, and $(\theta_{p}, \phi_p)$ are the coordinates of the pulsar. These coordinates are related to right ascension (RA) and declination (Dec) by: $\theta = \pi/2 - \mathrm{Dec}$ and $\phi = \mathrm{RA}$. $\alpha$ is the angular separation between them, and $\psi$ denotes the polarization angle of the CGW.

In Equation (\ref{eq:residual}), $s_A(t_p)$ and $s_A(t)$ denote timing residuals caused by CGW passing the pulsar and Earth, which are called the pulsar term and Earth term, respectively \citep{hellings1981spacecraft, jenet3c66b}. The pulsar time $t_p$ and Earth time $t$ satisfy the following relationship:
\begin{equation}
t_p = t - L_p(1 - \cos{\alpha})/c,
\end{equation}
where $L_p$ represents the pulsar's distance from the Earth and $c$ is the speed of light. In the Newtonian approximation, $s_A(t)$ takes the following form:
\begin{equation}
\begin{split}
s_+(t) &= \frac{\mathcal{M}^{5/3}}{d_L \omega(t)^{1/3}}[-\sin 2\Phi(t)(1+\cos^2\iota)], \\
s_\times(t) &= \frac{\mathcal{M}^{5/3}}{d_L \omega(t)^{1/3}}[2\cos 2\Phi(t)\cos\iota],
\end{split}
\end{equation}
where $\iota$ is the inclination angle of the SMBBH orbit with respect to the observer, $d_L$ is the luminosity distance to the source, and $\mathcal{M}$, called the ``chirp mass", is defined as $\mathcal{M}=(m_1m_2)^{3/5}/(m_1+m_2)^{1/5}$, with $m_1$ and $m_2$ being the masses of the two component black holes, $\omega$ is the angular orbital frequency. In Earth terms, if we set $t_0$ as a reference time, $\omega_{0}=\omega(t_{0}) = \pi f_\text{GW}$ where $f_\text{GW}$ is the gravitational wave frequency, and $\omega(t)$ and $\omega(t_0)$ satisfy the following relationship:
\begin{equation}
\omega(t) = \omega_{0} \left[1 - \frac{256}{5} \mathcal{M}^{5/3} \omega_{0}^{8/3} (t-t_{0})\right]^{-3/8}.
\label{eq:evolve}
\end{equation}
In this work, we set $t_0$ as the earliest observation time of PPTA DR3. Similarly, for the gravitational wave phase $\Phi$, if $\Phi_{0}$ is the initial Earth term phase,  $\Phi(t)$ and $\Phi_{0}$ satisfy the following relationship:
\begin{equation}
\Phi(t)=\Phi_{0} + \frac{1}{32} \mathcal{M}^{-5/3} [\omega_{0}^{-5/3}-\omega(t)^{-5/3}].
\label{eq:phi}
\end{equation}
For pulsar term signals, the phase is related to the initial pulsar phase $\Phi_i$ via equation (\ref{eq:phi}). We can see that the CGW signal is determined by the following eight parameters: $\theta$, $\phi$, $\psi$, $\iota$, $\Phi_0$, $f_\text{GW}$, $d_L$. Additionally, to include the pulsar term, two more parameters are required for each pulsar: the distance to the pulsar $L_{p}^{i}$, and the initial pulsar phase $\Phi_i$. Unless otherwise specified in the following sections, we will process the data using the full signal model. Finally, we can define the signal amplitude parameter:
\begin{equation}
h_0 = \frac{2 \mathcal{M}^{5/3} (\pi f_\text{GW})^{2/3}}{d_L}.
\label{eq:strain dl}
\end{equation}

\subsection{Bayesian methods} \label{subsec:baye_mothods}

The log likelihood function for pulsar timing residuals can be expressed as \citep[see, e.g.,][]{quickcw}:
\begin{equation}
\log L = -\frac{1}{2}(\delta t - s | \delta t - s) - \frac{1}{2}\log \det (2\pi C),
\end{equation}
where $(a|b)=a^T C^{-1} b$ and the noise covariance matrix $C=N+TBT^T$. Here, $N$ represents the white noise covariance matrix, and $T$ is the design matrix of the timing model, red noise, dispersion measure variations, and other noise processes, $B$ is the prior matrix for hyperparameters of noise models.

\begin{table}[h]
    \centering
    \begin{tabular}{ccc} 
        \toprule 
        Parameter           & Search & Upper Limits \\
        \midrule 
        $\cos \theta$       & Uniform ($-1$, 1)     & Uniform ($-1$, 1) \\
        $\phi$              & Uniform (0, 2$\pi$) & Uniform (0, 2$\pi$) \\
        $\log_{10} \mathcal{M}$     & Uniform (7, 10)     & Uniform (7, 10) \\
        $\log_{10} f_\text{GW}$ & Uniform ($-9$, $-8$)    & Constant (many) \\
        $\log_{10} h_0$       & Uniform ($-18$, $-11$)  & LinExp ($-18$, $-11$) \\
        $\Phi_0$            & Uniform (0, $2\pi$) & Uniform (0, $2\pi$) \\
        $\psi$              & Uniform (0, $\pi$)  & Uniform (0, $\pi$) \\
        $\cos \iota$        & Uniform ($-1$, 1)     & Uniform ($-1$, 1) \\
        $\Phi_i$            & Uniform (0, $2\pi$)  & Uniform (0, $2\pi$) \\
        $L_{p}^{i}$               & Normal  $\mathcal N$($L_i$, $\sigma_i$) & Normal  $\mathcal N$($L_i$, $\sigma_i$) \\
        \bottomrule 
    \end{tabular}
    \caption{List of parameters of the CGW model with their respective priors and ranges. Most priors are the same for the search and upper limits, except that 1) we adopt a uniform distribution for $h_0$, which is more conservative while setting upper limits; 2) we place upper limits in a number of frequency bins.}
    \label{tab:prior}
\end{table}

Table \ref{tab:prior} shows the priors used for the CGW parameters. For most parameters, we adopt either a uniform distribution or a log-uniform distribution as priors. Measuring celestial distances has always been a challenging problem in astronomy, making it difficult to accurately describe the pulsar term. In this paper, we treat the pulsar distance as a free parameter \citep{eptadr2cw, aggarwal2019nanograv} that needs to be estimated based on the data. We set the prior for $L_{p}^{i}$ as a Gaussian distribution $\mathcal{N}(L_i, \sigma_i)$; in Table \ref{tab:distance} of Appendix \ref{app:psr_distance}, we list the distances and uncertainties of PPTA pulsars.

We use the {\tt\string enterprise} \citep{enterprise} and {\tt\string enterprise\_extensions} \citep{enterprise_extensions} packages to perform Bayesian inference. Our goal is to search for evidence of CGWs and to determine the posterior distribution of CGW parameters. The parameter space is explored using the Markov Chain Monte Carlo (MCMC) sampler package {\tt\string PTMCMCSampler} \citep{justin_ellis_2017_1037579}.

To evaluate the statistical support for CGW evidence, we calculate the Bayes factor (BF) using product-space sampling \citep{model_choice_carlin1995, model_choice_godsill2001}. This method was first applied to PTA gravitational-wave searches by \citet{ng11yr_gwb}, with formal derivation in the PTA context presented in \citet{Taylor2020-bumpy}. Our implementation uses the {\tt\string enterprise} package. It involves creating a hypermodel that includes all the parameters of the two models being compared, and, additionally, a parameter to track the migration of Markov chains. The Bayes factor is determined by comparing the ratio of samples from the two model chains. For example, if we consider two models $A$ and $B$, then the BF is the ratio $n_A$/$n_B$, where $n_A$ and $n_B$ denote the number of samples in the chain corresponding to the models $A$ and $B$, respectively. In the case of no significant support for the presence of CGWs in our data, we calculate the 95\% credibility upper limits on the signal amplitude $h_0$, which is defined as the 95\% quantile of themarginalized posterior distribution.

\subsection{Frequentist methods} \label{subsec:freq_methods}
We also perform a search for CGWs using the frequentist approach described in \citet{Ellis_Frequentist_2012}, including both the $\mathcal{F}_\text{e}$ and $\mathcal{F}_\text{p}$ statistic as implemented in the \texttt{enterprise\_extensions} software package. 

The $\mathcal{F}_\text{e}$ statistic is a function of the GW frequency as well as the source sky position. It is the coherent $\mathcal{F}$-statistic \citep{Ellis_Frequentist_2012}, which includes only the Earth term and treats the pulsar term as a noise source; it is applicable when the pulsar-term frequency evolution exceeds the size of a frequency resolution bin ($1/T_{\rm obs}$).
The false alarm probability (FAP), $P(\mathcal{F} > \mathcal{F}_0)$, for the $\mathcal{F}_\text{e}$ statistic is calculated according to the methodology defined in \citet{Babak_EPTA_CW_2015}. Given a threshold, $\mathcal{F}_0$, one then takes the integral of the probability distribution in the form\\
\begin{equation}
    P(\mathcal{F}_\text{e} > \mathcal{F}_0) = \int_{\mathcal{F}_0}^{\infty} p_0(\mathcal{F}_\text{e}) d\mathcal{F}_\text{e} = (1+\mathcal{F}_0)e^{-\mathcal{F}_0},
\end{equation}
where $p_0(\mathcal{F}_e)$ is the background distribution of $\mathcal{F}_e$ in the data without signals. 

The $\mathcal{F}_\text{p}$ statistic is more sensitive to the sources where frequency evolution is negligible. The $\mathcal{F}_\text{p}$ statistic is the incoherent $\mathcal{F}$-statistic and includes the pulsar term in the signal model. If the frequency evolution of the source is low enough that the Earth term and the pulsar term are the same, then the signal is a sum of two sinusoids which are of different phase \citep{Ellis_Frequentist_2012}. A FAP associated with the $\mathcal{F}_\text{p}$ statistic is defined in a similar way to that of the $\mathcal{F}_\text{e}$ statistic, \\
\begin{equation}
P(\mathcal{F}_p > \mathcal{F}_0)=\int_{\mathcal{F}_{0}}^{\infty}p_{0}(\mathcal{F}_p)d\mathcal{F}
=\exp(-\mathcal{F}_{0})\sum_{k=0}^{n/2-1}\frac{\mathcal{F}_{0}^{k}}{k!}.
\end{equation}
In both the case of the $\mathcal{F}_\text{e}$ and the $\mathcal{F}_\text{p}$ statistic, the number of trials should be taken into account. In the case of the $\mathcal{F}_\text{e}$ statistic, the number of trials is represented by the number of frequencies searched over, as well as the number of different sky positions. In the case of the $\mathcal{F}_\text{p}$ statistic, only the number of frequencies is relevant, as this statistic is not sensitive to sky position. Therefore, the global FAP is given by

\begin{equation}
\label{eq:fap}
P_\text{t}(\mathcal{F}_{0})=1-\left[ 1-P(\mathcal{F} > \mathcal{F}_{0}) \right]^{N_{\text{t}}},
\end{equation}
where $N_\text{t}$ is the number of trials.
The threshold is set as a FAP of $10^{-4}$, as in \citet{Ellis_Frequentist_2012}. If the statistic at a specific frequency has an associated FAP below this threshold, a detection can be claimed.

\section{Results} \label{sec:result}

\subsection{Bayesian search} \label{subsec:search}

\begin{figure}[h]
    \centering
    \includegraphics[width=\linewidth]{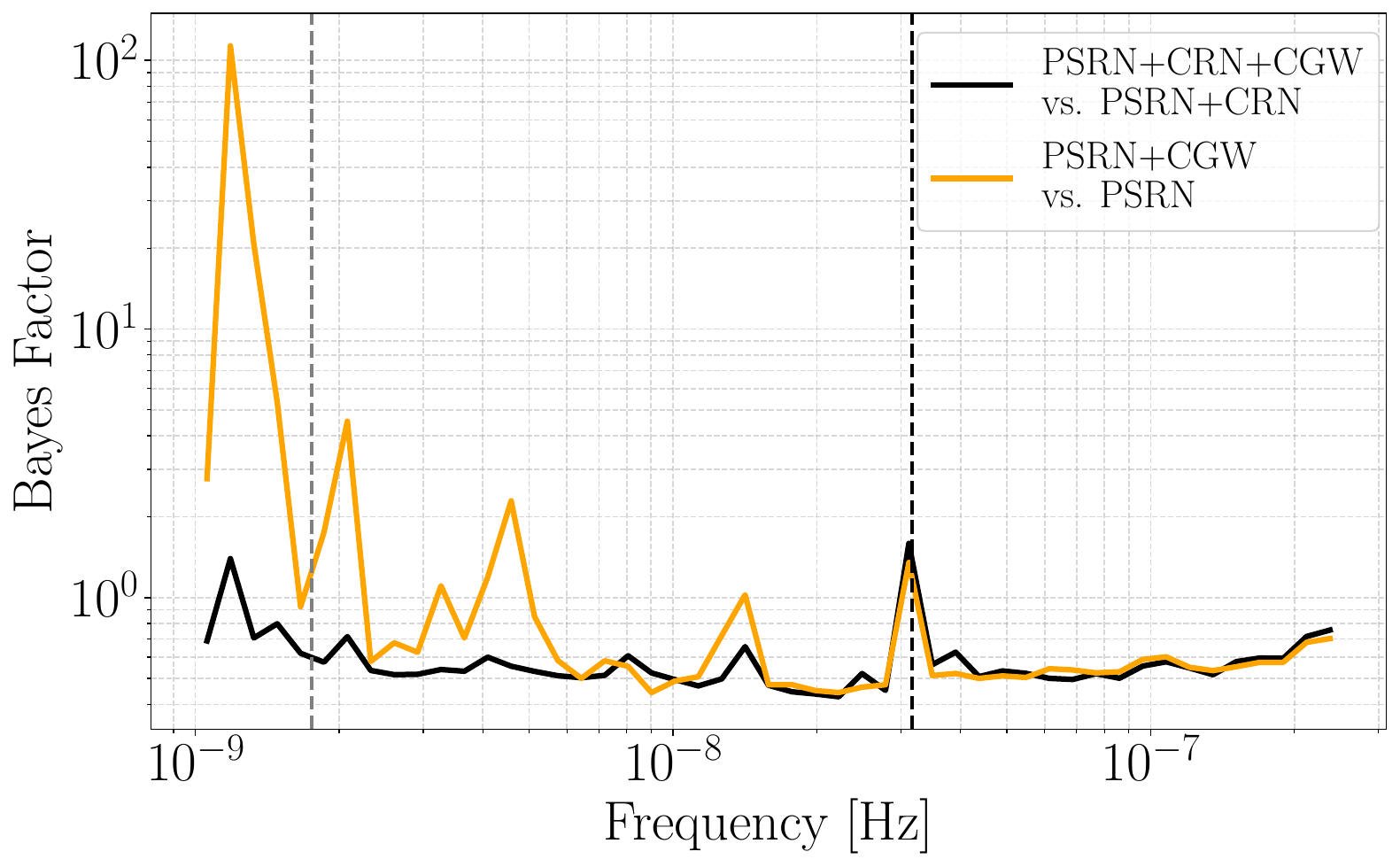}
    \caption{Savage-Dickey Bayes factors in support for a CGW at each GW frequency. The orange curve indicates results for the case where common red noise (CRN) is not included in the analysis, while the black curve is for the case where it is included. We found that no CGW are detected in the PPTA DR3 if a CRN is accounted for. The black and gray vertical dotted lines mark, respectively, the frequency of $1/\mathrm{yr}$ and of $1/T_{\mathrm{obs}}$. These two vertical lines retain the same meaning in all subsequent figures of this paper.}
    \label{fig:S-D_BF}
\end{figure}

In Figure \ref{fig:S-D_BF}, we plot the BFs in favor of the presence of a CGW for 50 logarithmically spaced frequency bins from $10^{-9}$ to $10^{-6.8}$ Hz. The BFs are calculated for two cases depending on whether or not a CRN is included in our analysis, namely, PSRN+CGW against PSRN (orange curve), and PSRN+CRN+CGW against PSRN+CRN (black curve). We notice that multiple peaks in the BF plot are present for the pulsar noise only case, and these peaks disappear after accounting for the CRN (except a small peak at $\unit[1] {{\rm yr}^{-1}}$). These peaks are located at low frequencies where a CRN signal contributes most power. It is also worth mentioning that the orange curve exhibits a small peak around 4–5 nHz, with a BF of approximately 2.3. Both EPTA and NANOGrav reported a suspicious signal in this frequency range \citep{agazie2023nanogravcw, eptadr2cw}; they found a BF of $\sim 4$ in favor of CGW on top of PSRN+CRN. Note that the BFs for such a signal were reduced to the order of 1 while the HD correlation was included in their analyses.
Because the support for HD correlation in the Bayesian analysis of PPTA DR3 is marginal, with a BF of 1.5 \citep[see,][for details]{reardon2023search}, we do not include the HD correlation in our search for CGW signals; see Appendix \ref{app:posterior} for additional comments.

\begin{figure}[h]
    \centering
    \includegraphics[width=\linewidth]{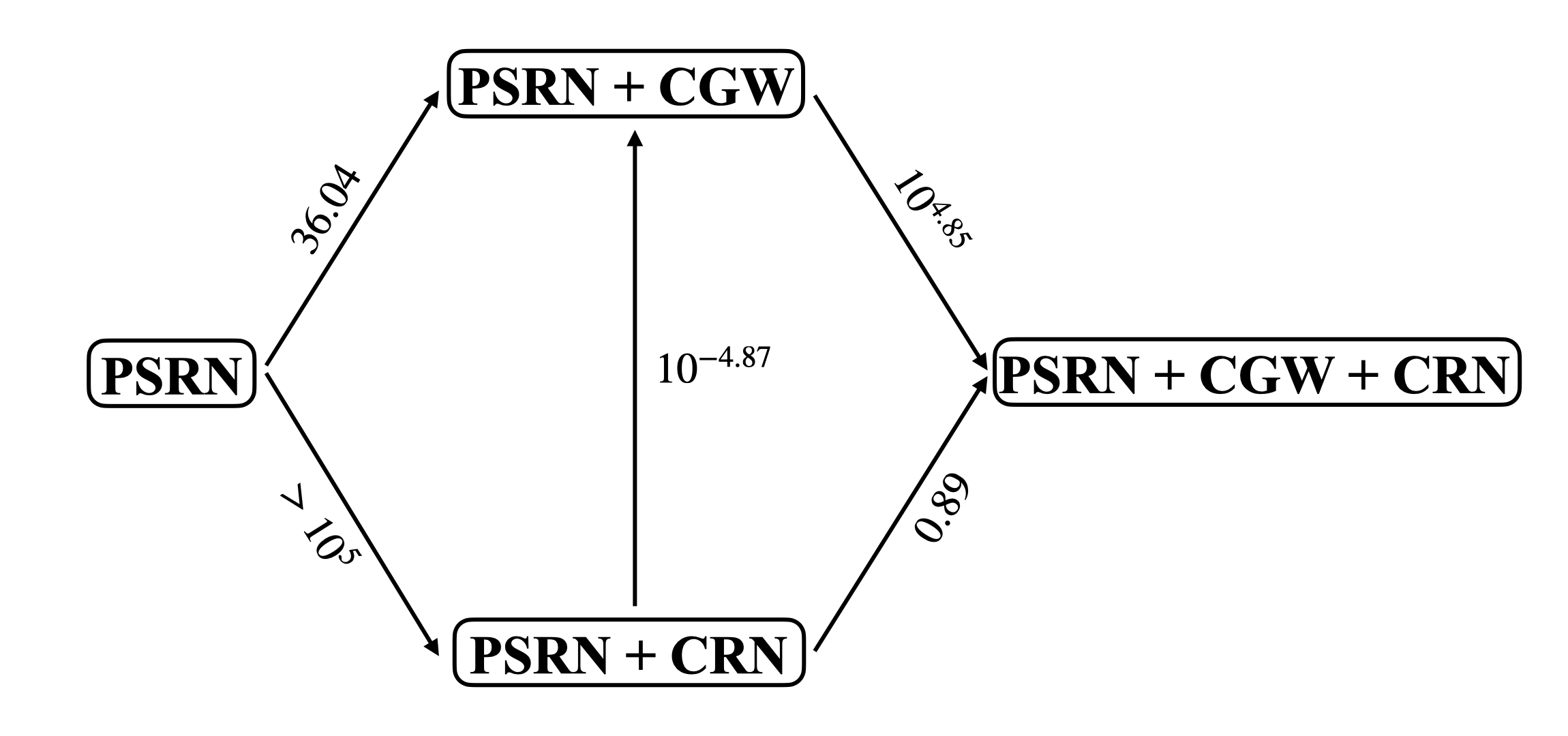}
    \caption{Bayes factor for comparison between different models, where numerical values represent the extent to which the data favors the model to which the arrow points. CGW here refers to a continuous wave signal with frequencies between 1 and $\unit[2.5]{nHz}$, the frequency range where the BF peaks in Figure \ref{fig:S-D_BF}.}
    \label{fig:model_selection}
\end{figure}

To further quantify statistical support for a CRN and CGW, we calculate BFs in the frequency range between 1 and $\unit[2.5]{nHz}$ where the two highest peaks in the orange curve of Figure \ref{fig:S-D_BF} are found. The results are shown in Figure \ref{fig:model_selection}.
When using the PSRN model as the null hypothesis, a strong CRN signal is evident in the data, with the Bayesian factor reaching $10^{5}$, consistent with the findings of \citet{reardon2023search}. Comparing  PSRN+CGW with PSRN+CRN, we find a BF of $10^{-4.87}$, indicating that the signal is more likely to be interpreted as CRN rather than CGW. Considering the PSRN+CRN+CGW model with the PSRN+CRN model, the BF is merely 0.89, and therefore we conclude the search with no evidence of CGW in the data.
In appendices \ref{app:posterior} and \ref{app:CW_ON_CRN}, we present the posterior distributions of CGW parameters and the CRN parameters for models with different signal+noise combinations.

\subsection{Frequentist Search} \label{subsec:freq_search}

\begin{figure}[h]
    \centering

    \begin{subfigure}{\linewidth}
        \centering
        \includegraphics[width=\linewidth]{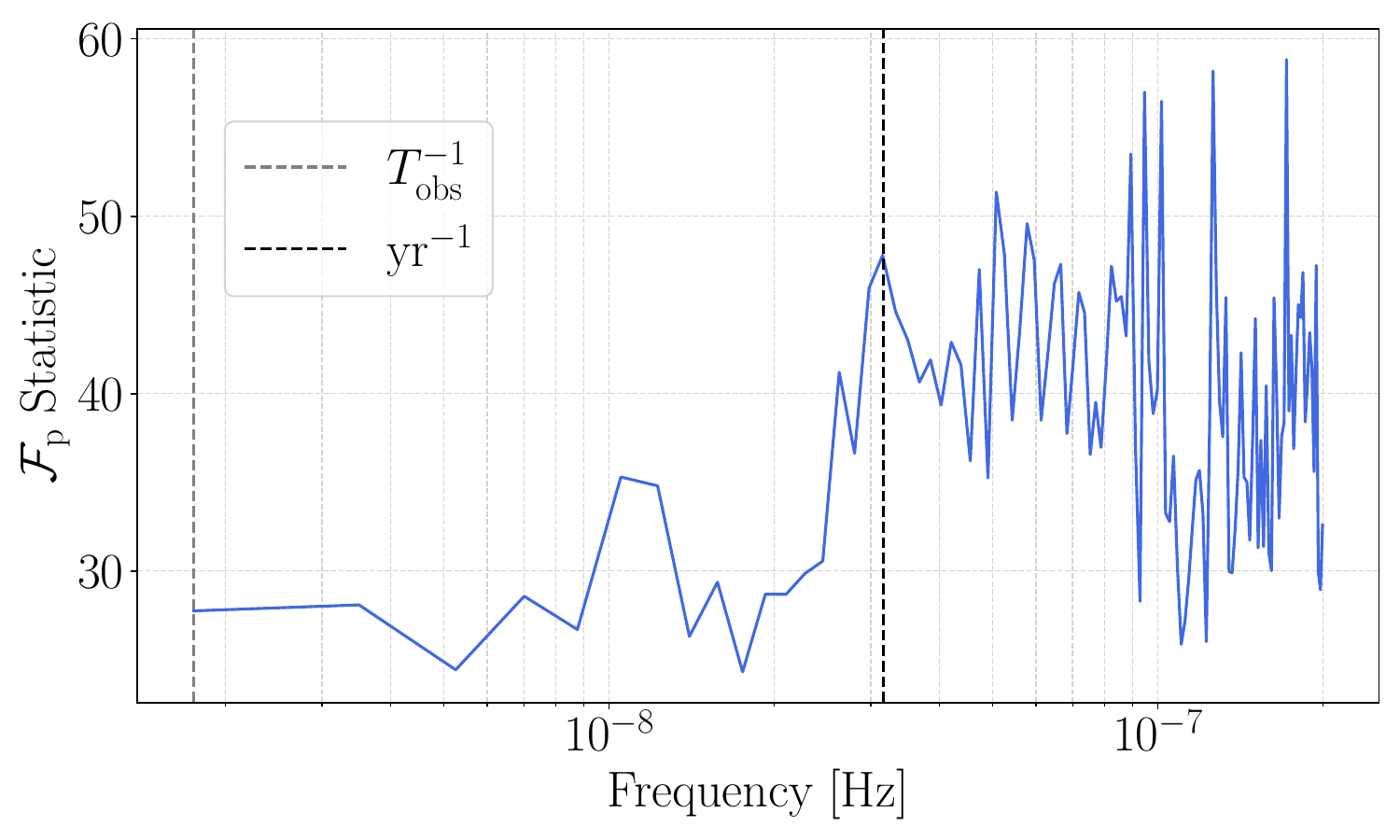}
        \caption{~}
        \label{fig:fp_stat}
    \end{subfigure}

    \vspace{1em} 

    \begin{subfigure}{\linewidth}
        \centering
        \includegraphics[width=\linewidth]{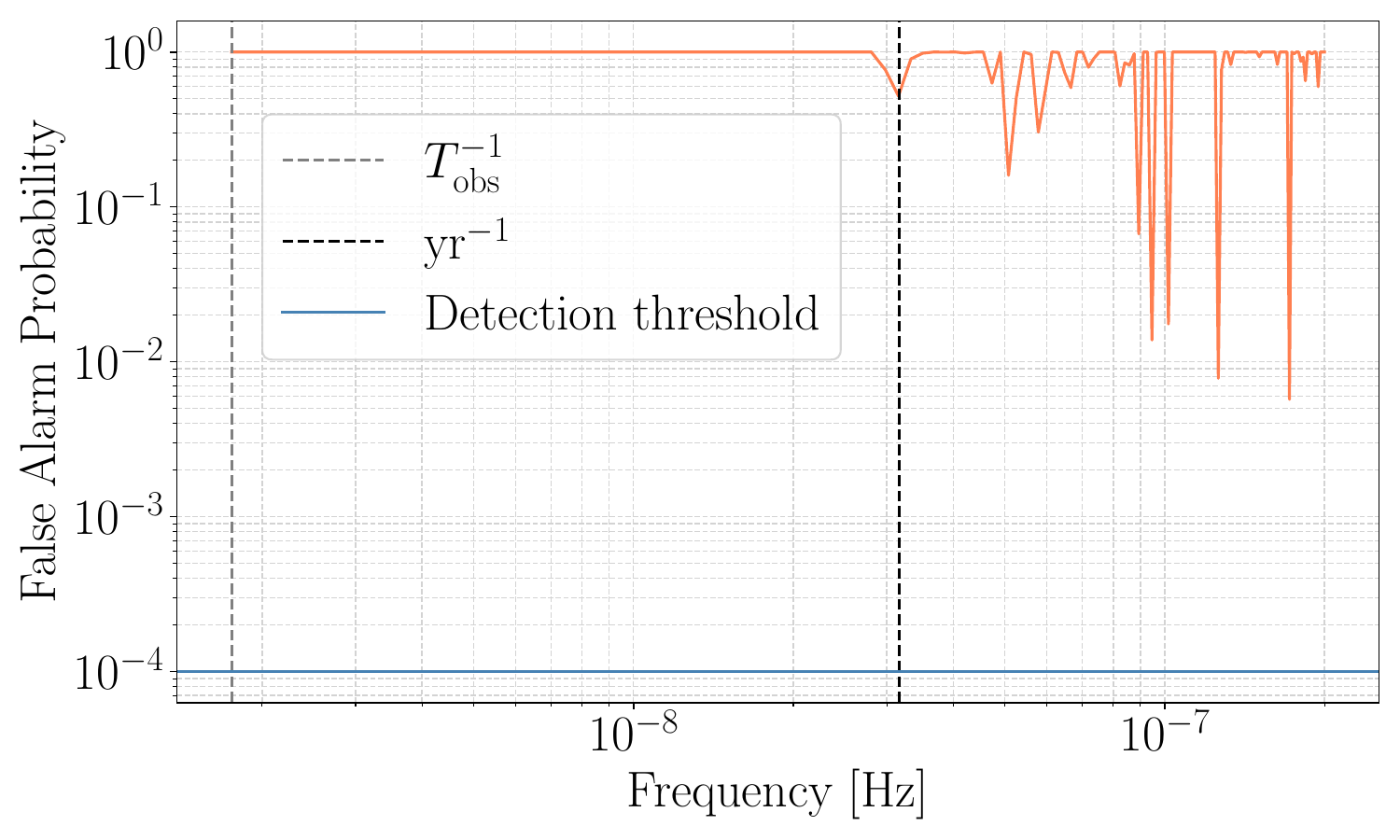}
        \caption{~}
        \label{fig:fp_fap}
    \end{subfigure}

    \caption{The values of the all-sky CGW frequentist detection statistic $\mathcal{F}_\text{p}$ and the respective FAP are shown against GW frequency.}
    \label{fig:fp_stat_combined}
\end{figure}

We perform a frequentist search for both the Earth term and the Pulsar term of the CGW by computing the $\mathcal{F}_\text{p}$ statistic in the frequency range from $T^{-1}_{\text{obs}}$ to $2 \times 10^{-7}$~Hz. 
We perform a search simultaneously with the modeling of CRN and PPTA noise terms found in~\citet{reardon2023noises}. For simplicity, each $\mathcal{F}$ statistic value is calculated based on assuming maximum-\textit{a-posteriori} values of noise parameters from single-pulsar analyses using marginalized distributions for every parameter. 
For a common power-law signal, we assume the values from~\citet{reardon2023search}.
The values of the $\mathcal{F}_\text{p}$ statistic at each frequency are shown in Figure~\ref{fig:fp_stat}. 
The respective values of FAP are shown in Figure~\ref{fig:fp_fap}.
We find the maximum value of $\mathcal{F}_\text{p}= 58.84$ at $1.72 \times 10^{-7}$ Hz, corresponding to $\text{FAP}= 5.72 \times 10^{-3}$. 
The second highest value of $\mathcal{F}_\text{p}= 58.19$ is at $1.26 \times 10^7$ Hz, corresponding to $\text{FAP}=7.82 \times 10^{-3}$. 
The third highest value of $\mathcal{F}_\text{p}= 57.00$ is at $9.46 \times 10^{-8}$ Hz, corresponding to $\text{FAP}= 1.38 \times 10^{-2}$. 
We do not find a value of $\mathcal{F}_\text{p}$ with FAP below the detection threshold of $10^{-4}$. Unlike in Section~\ref{subsec:search} which describes the results of the Bayesian search, here we find the most statistically significant outliers at high frequencies above $\unit[1]{\text{yr}^{-1}}$.

\begin{figure}[h]
     \centering
     \includegraphics[width=\linewidth]{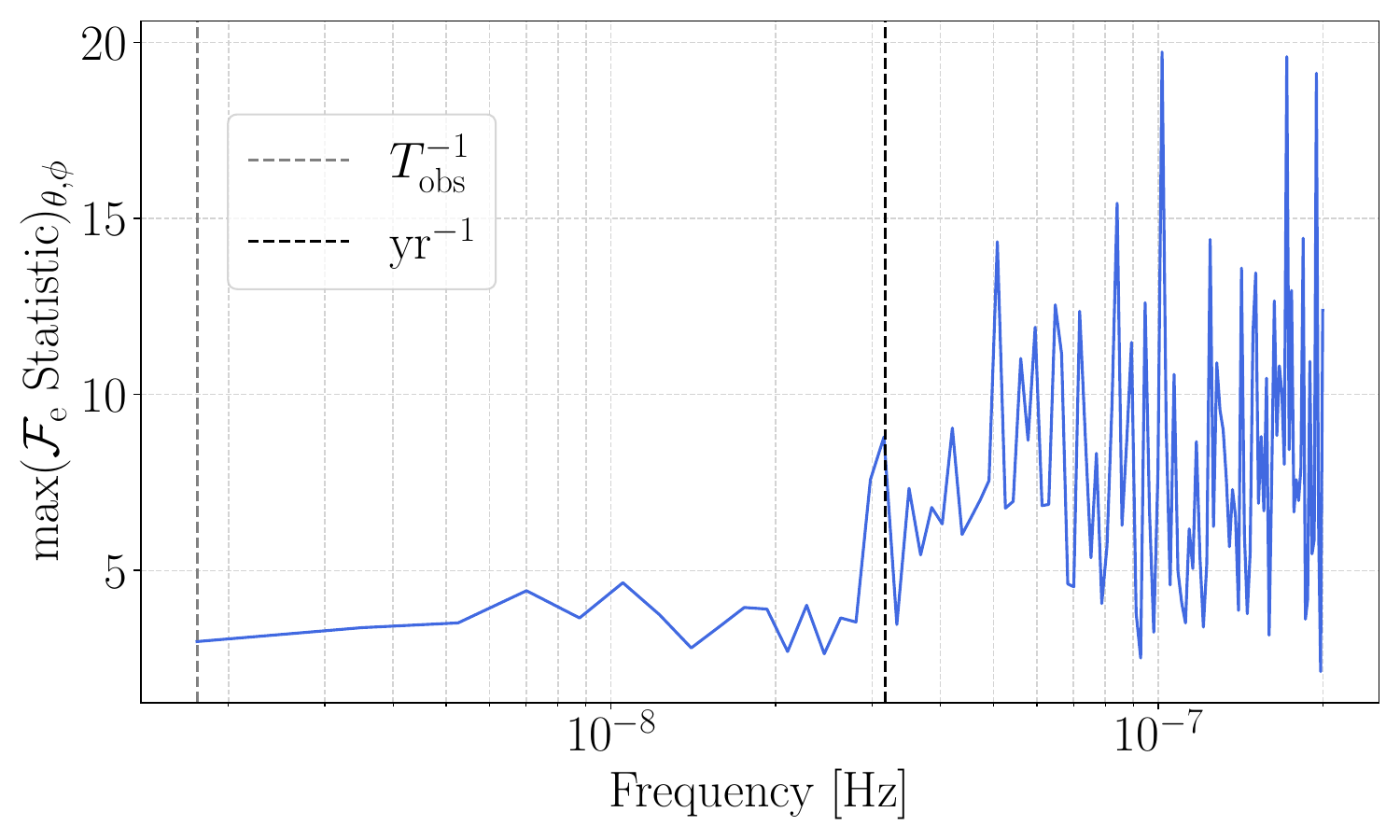}
     \caption{The values of the CGW detection statistic $\mathcal{F}_\text{e}$, maximized over a range of $192$ uniformly-distributed sky positions for every GW frequency.}
     \label{fig:Fe_stat_results}
 \end{figure}

Next, we perform the frequentist search for only the Earth term of the CGW by computing the $\mathcal{F}_\text{e}$ statistic for 192 sections of the sky which is determined by $N_{\textrm{pix}} = 12 N_{\textrm{side}}^2$ \citep{gorski2005healpix} where we choose $N_{\textrm{side}} = 4$.
The maximum value $\mathcal{F}_\text{e}$ of each sky position for each frequency is plotted in Figure~\ref{fig:Fe_stat_results}. 
We find the highest $\mathcal{F}_\text{e}= 19.72$ at $1.02 \times 10^{-7}$ Hz.
We find this values corresponds to $\text{FAP}=1$ due to the high number of trials.
This presents a conservative result because not all $\mathcal{F}_\text{e}$ statistic values are independent at trialed sky positions.
Based on this result, we conclude that the frequentist search does not show significant evidence for a CGW. 

\subsection{Upper limits} \label{subsec:limit}
\begin{figure}[t]
    \centering
    \includegraphics[width=\linewidth]{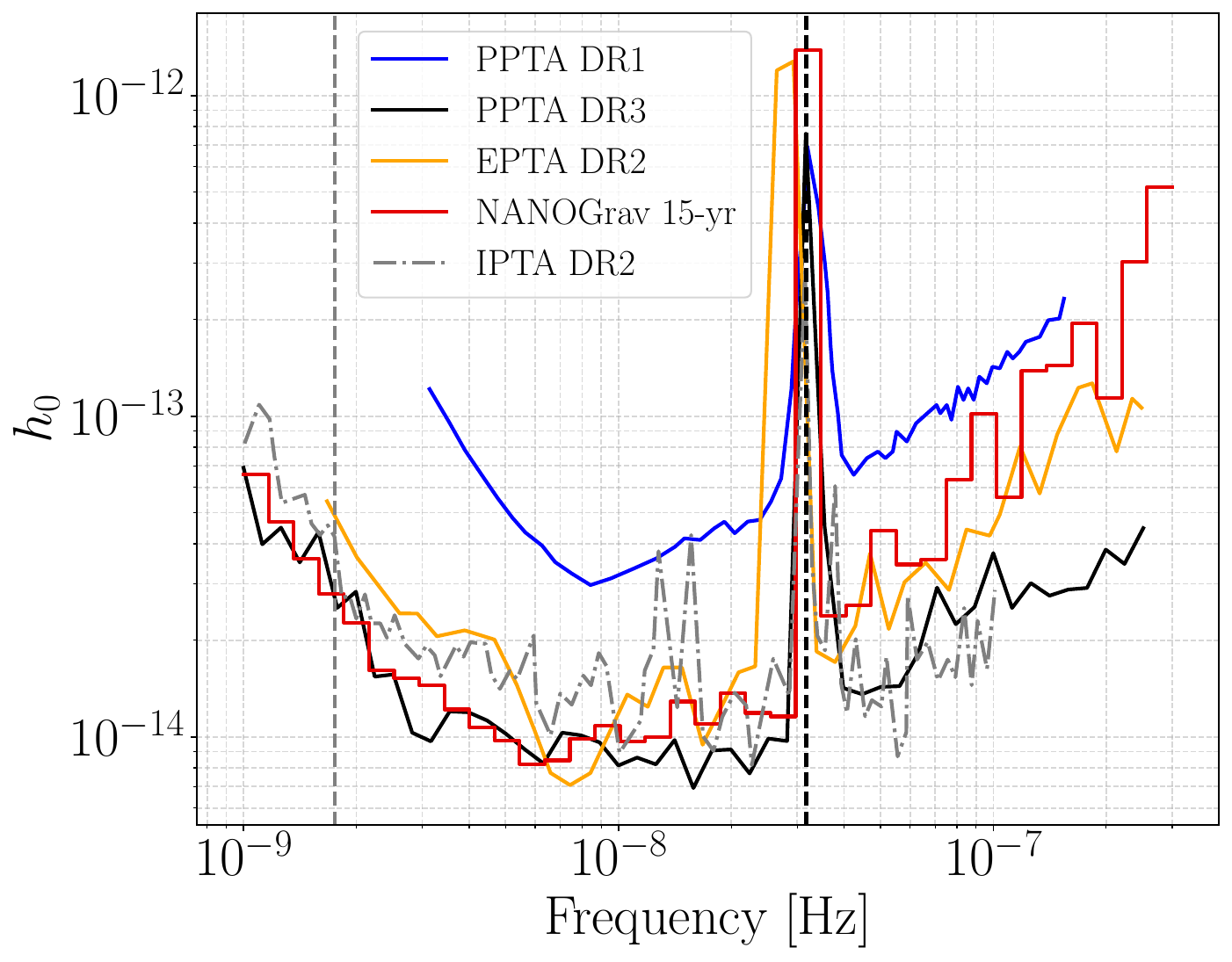}
    \caption{All-sky 95\% credibility upper limits on the CGW strain amplitude $h_0$. The PPTA DR3 (black) is more sensitive than the DR1 (blue) by at least a fractor of four over a wide frequency range. For comparison, the most recent strain upper limits from the EPTA, NANOGrav, and IPTA are also included in the figure.}
    \label{fig:UL}
\end{figure}

Since we find no evidence for the existence of CGW in the PPTA DR3 data set, we place upper limits on the CGW strain amplitude. In Figure \ref{fig:UL}, we plot the 95\% credibility upper limits on $h_0$ (in black) for 50 frequency bins log-uniformly distributed between 1.6 nHz and 250 nHz. In the frequency range from 3 nHz to 25 nHz, we constrain $h_0 < 1 \times 10^{-14}$, with the lowest limit reaching $7 \times 10^{-15}$. For comparison, we also plot the upper limits (in blue) derived from PPTA DR1 in the figure. It can be seen that, in comparison to DR1, our DR3 data set has improved by at least a factor of four, and the accessible frequency range has been significantly extended. These improvements are mainly due to the longer data span (i.e., 6 years increased to 18 years) and larger pulsar array (i.e., 20 pulsars to 30 pulsars) of the data set. Additionally, we notice that the upper limit curve of DR1 is smoother. This is partly because PPTA DR1 results were derived using the frequentist method \citep{zhu2014all}; it could also be due to differences in noise models, with PPTA DR3 employing a more complex noise model.
Also plotted in Figure \ref{fig:UL} are the upper limits derived with the IPTA DR2 \citep{Falxa23-ipta-dr2cw}, EPTA DR2 \citep{eptadr2cw}, and NANOGrav 15-yr data set \citep{agazie2023nanogravcw}.
It can be seen that the PPTA DR3 limits are comparable to contemporary results; they are notably better than EPTA and NANOGrav limits at frequencies above $\unit[1]{\text{yr}^{-1}}$.
This is probably due to higher effective observing cadence for the PPTA data set.
For the same reason, the IPTA DR2 upper limits above $\unit[1]{\text{yr}^{-1}}$ are also better than EPTA and NANOGrav limits.

\begin{figure}[h]
    \centering
    \includegraphics[width=\linewidth]{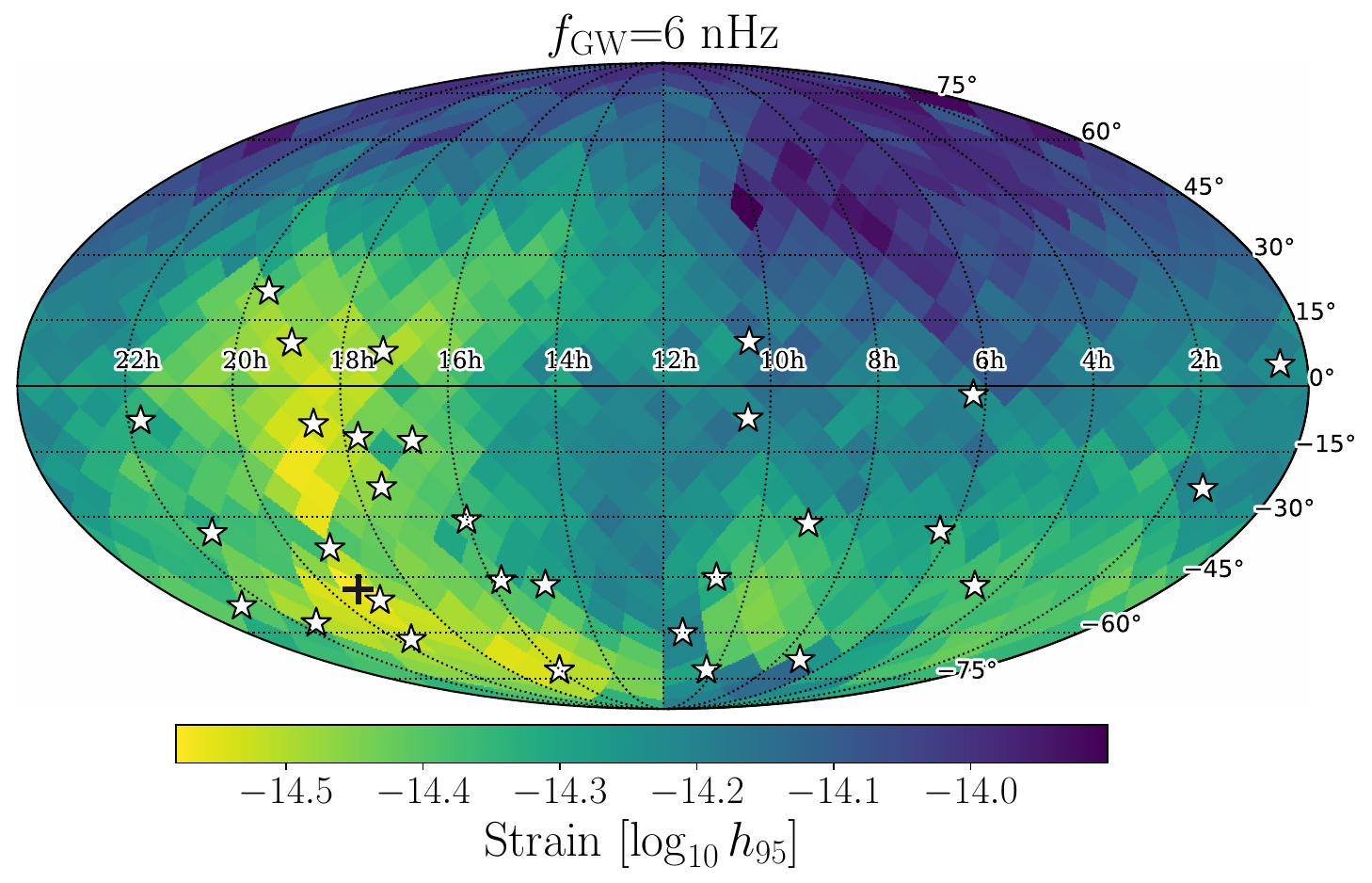}
    \caption{Sky map of 95\% credibility upper limits on the CGW signal amplitude $h_0$ at 6 nHz for the PPTA DR3. Pulsars are marked as white stars, while the black ``+" symbol indicates the most sensitive sky location.}
    \label{fig:skymap_strain}
\end{figure}

In Figure \ref{fig:skymap_strain}, we plot the GW strain upper limits at 6 nHz as a function of sky location. The white stars indicate the sky locations of the PPTA pulsars and the black `+' indicates the most sensitive location. As expected, the region of the sky most sensitive to CGWs corresponds to where most PPTA pulsars are located. At the most sensitive pixel, the upper limit is $2.6 \times 10^{-15}$, while at the least sensitive pixel, it is $1.3 \times 10^{-14}$, showing a sensitivity variation by a factor of 5.

\begin{figure}[h]
    \centering
    \includegraphics[width=\linewidth]{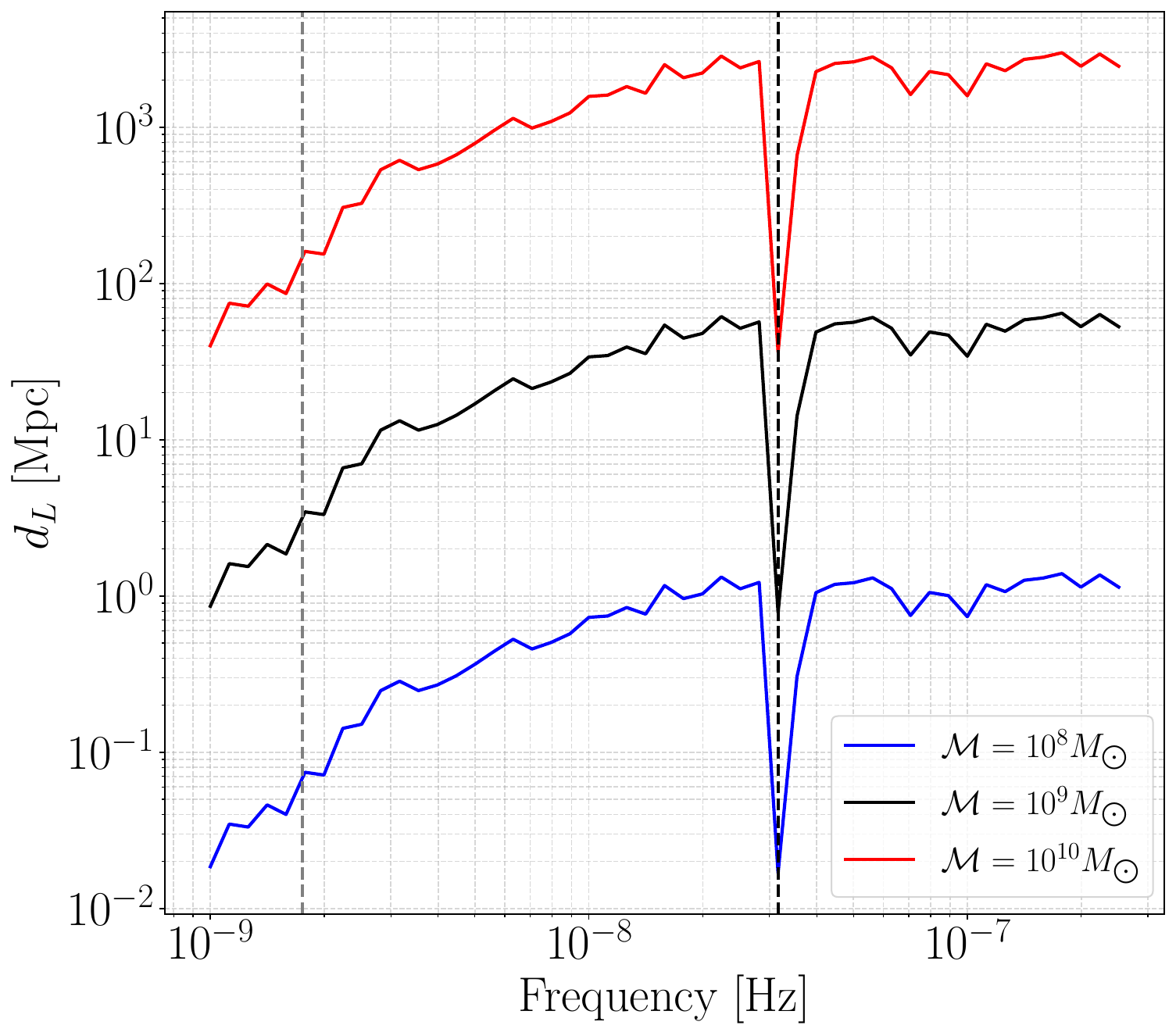}
    \caption{Horizon luminosity distance, $d_L$, obtained from the sky averaged 95\% upper limit on strain amplitude $h_0$ shown in Figure \ref{fig:UL}. The luminosity distance limit is calculated with Equation (\ref{eq:strain dl}) for three chirp masses: $10^8M_{\odot}$, $10^9M_{\odot}$ and $10^{10}M_{\odot}$.}
    \label{fig:DL_limit}
\end{figure}

Our upper limits on $h_0$ can be converted to lower limits on the luminosity distance to the source for a given chirp mass using Equation (\ref{eq:strain dl}). In Figure \ref{fig:DL_limit}, we plot the luminosity distance limits as a function of frequency for three chirp mass values: $\mathcal{M}=[10^8M_{\odot}, 10^9M_{\odot}, 10^{10}M_{\odot}]$. Because of the scaling relation of $h_0 \sim f_{\rm GW}^{2/3}$, we can see that the luminosity distance limits plateau after reaching the peak at around 20 nHz (except a dip at $1/{\rm yr}$). For a chirp mass of $10^9M_{\odot}$, the luminosity distance lower limit is above 20 Mpc for a frequency above 6 nHz. To place this in astrophysical context, the closest galaxy clusters, the Virgo and Fornax cluster are located at 16.5 Mpc and 19 Mpc, respectively.

\begin{figure}[h]
    \centering
    \includegraphics[width=\linewidth]{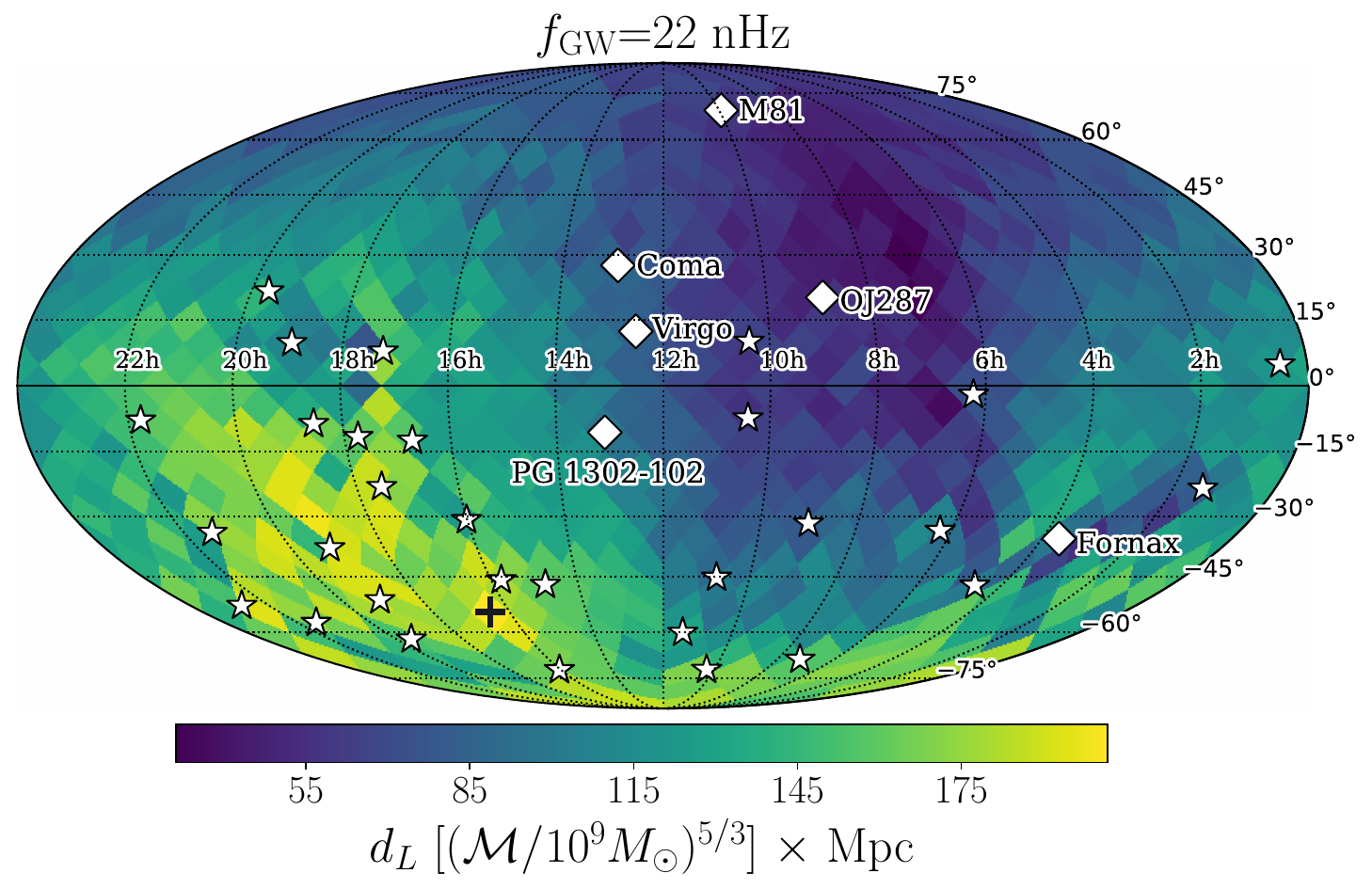}
    \caption{Sky map of the 95\% lower limit on the luminosity distance to individual SMBBHs with $\mathcal{M}=10^9 M_{\odot}$ and $f_\text{GW}=22$~nHz. White squares indicate the positions of known SMBBH candidates or galaxy clusters that could host a SMBBH.}
    \label{fig:dl_skymap}
\end{figure}

Figure \ref{fig:dl_skymap} illustrates the distance up to which a circular SMBBH would produce a detectable signal at a frequency of 22 nHz in our data set. The white stars indicate the locations of the PPTA pulsars, and white diamonds mark the positions of several well-known SMBBH candidates and a few nearby galaxy clusters: Virgo, Fornax and Coma. OJ287 is one of the most well-monitored blazars in the X-ray–UV–optical band with an orbital period of about 12 years \citep{komossa2023momo, sundelius1997numerical}. PG1302-102 is a quasar exhibiting a strong and smooth periodic optical signal that is interpreted as arising from a sub-parsec binary system of two supermassive black holes \citep{PG1302}; note that the periodicity detection was questioned in \citet{pg1302_102} and it was shown that a quasi-periodic oscillation model is preferred for its optical light curve \citep{ZhuThrane20}. M81 is a nearby galaxy showing evidence of a centi-parsec SMBBH with periodic jet wobbling and X-ray outbursts \citep{M81}.
Note that all the potential SMBBH candidates and nearby clusters are located in the less sensitive sky region, which emphasizes the importance of adding new well-timed pulsars in the other part of the sky to extend the astrophysical reach of the PPTA. The most straightforward way to achieve this is through the combination of regional PTA data sets via the IPTA collaboration.

\begin{figure}[h]
    \centering
    \includegraphics[width=\linewidth]{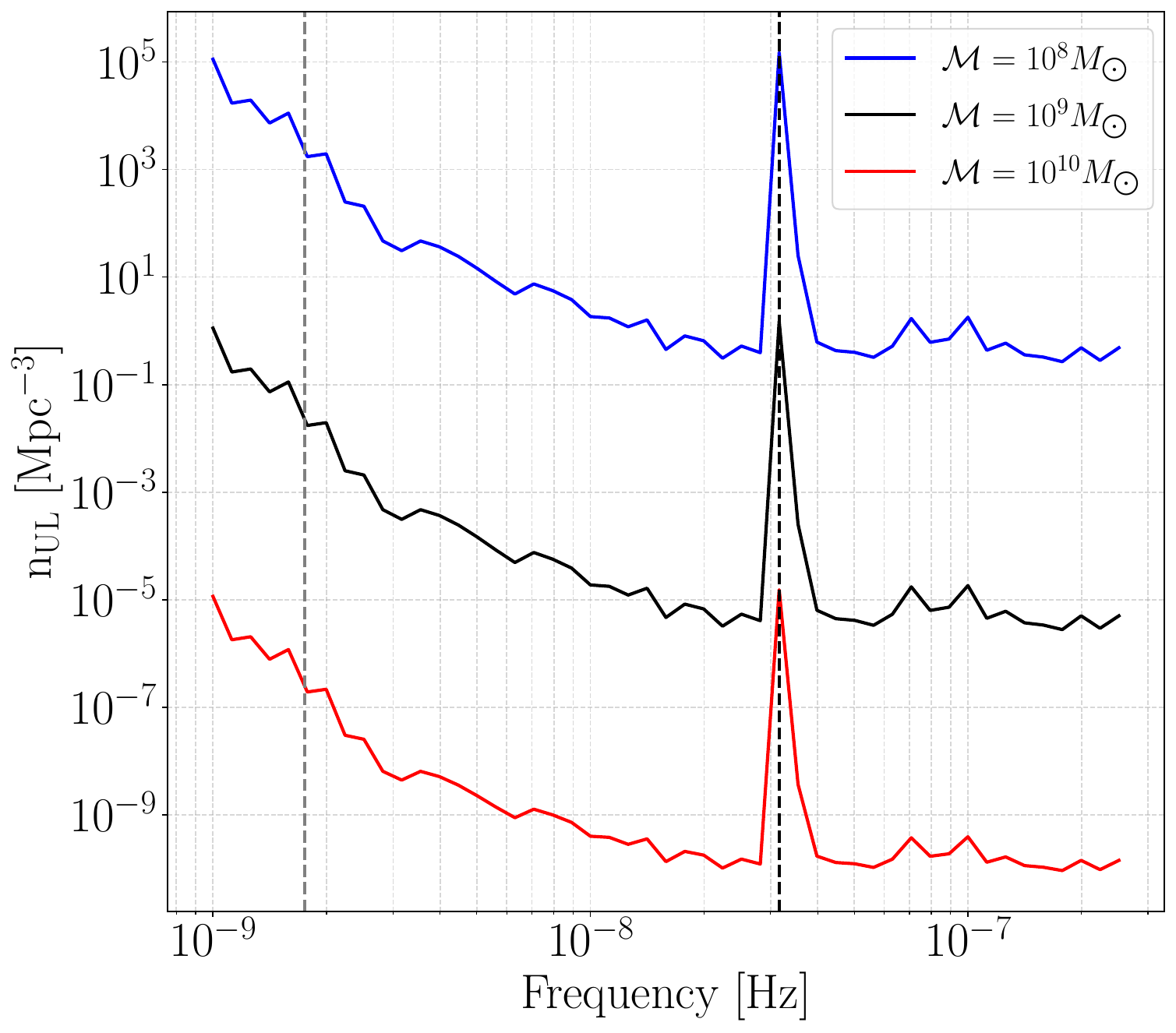}
    \caption{Upper limit on the number density ($n_\text{UL}$) of SMBBHs in the local universe as a function of GW frequency.  Different curves correspond to binaries with different chirp masses.  The density is measured in terms of comoving volume.}
    \label{fig:num_density}
\end{figure}

Based on our limits on the luminosity distance of SMBBHs, we can provide an upper limit on its local number density. Following \citet{ng12cw}, we assume that these sources follow compound Poisson distributions. For the number of SMBBH events, $\Lambda = n_c V_c$, the probability of no detection is given by $P_0 = e^{-\Lambda}$. Here $n_c$ represents the number density of SMBBH, and $V_c$ represents the volume of the sphere with the radius of the comoving distance $d_c$. Therefore, it is only necessary to find the upper limit $\Lambda_\text{UL}$ of the occurrence rate to satisfy $\int_{\Lambda_\text{UL}}^{\infty} e^{-\Lambda} d\Lambda = 1 - p_0$, to obtain the solution:

\begin{equation}
n_\text{UL}=\frac{-\ln{(1-p_0)}}{V_c},
\label{eq:num density}
\end{equation}
here $p_0$ is our detection threshold, which we set to 0.95. The relationship between co-moving distance $d_c$, luminosity distance $d_L$ and redshift $z$ is given by $d_c=d_L/(1+z)$, where $z$ is calculated for the relevant luminosity distance values using the {\tt astropy} package.
Figure \ref{fig:num_density} shows our results for the upper limit of SMBBH number density. As expected, the higher the chirp mass of a binary, the lower the limit we can provide.

Following \citet{zhu2014all}, we can also place constraints on the SMBBH merger rate using PPTA DR3. Specifically, if the SMBBH merger rate is $R$, the total number of merger events, $\mu$, can be expressed as $\mu = R\sum_{i=1}^{n} \epsilon V_i \Delta T(f_i)$. Here, $\epsilon$ represents the detection efficiency, set to 0.95 in this analysis. The term $V_i=4\pi d_{L,i}^3/3$ denotes the sensitive volume at frequency $f_i$, with both $V_i$ and $f_i$ derived from Figure~\ref{fig:DL_limit}. The quantity $\Delta T(f_i)$ is the duration a binary system spends in the $i$-th frequency bin, calculated using Equation~\ref{eq:evolve}. Assuming Poisson-distributed events, the probability of detecting no events is given by $e^{-\mu}$. Consequently, the frequentist 95\% confidence upper limit on the merger rate is $R_{95}=-\ln(1-0.95)/\sum_{i=1}^{n} \epsilon V_i \Delta T(f_i)$. 
Based on the results obtained from Figure~\ref{fig:DL_limit}, we determine the SMBBH merger rate upper limit to be $R_{95}=4.4\times10^{-5} (10^{10} M_{\bigodot}/\mathcal{M})^{10/3}$ Mpc$^{-3}$Gyr$^{-1}$.

\section{CONCLUSIONS} \label{sec:conclusions}

Global PTA collaborations have found a nHz signal that is consistent with a stochastic gravitational wave background.
This could be the highly anticipated signal produced by the cosmic population of SMBBHs, although early-Universe processes such as phase transitions and domain walls are possible origins as well.
One way that helps determine the origin of the nHz stochastic signal is to find a single SMBBH system through the detection of continuous gravitational waves.
In this paper, we use both Bayesian and frequentist methods to search for continuous wave signals in the PPTA DR3 data set. No evidence of such signals was found in our data, and we demonstrate that at least a factor of four improvement in detection sensitivity is achieved compared to the last continuous wave search with PPTA DR1.

We present both sky-averaged and directional upper limits on the intrinsic signal amplitude $h_0$. For example, at 6 nHz, our analysis excludes the presence of signals from any sky directions with $h_0 >7.4 \times 10^{-15}$ with 95\% confidence. At the same frequency bin, the upper limits on $h_0$ vary by a factor of 5 across the sky: in the most sensitive sky region, the signal amplitude cannot exceed $2.6 \times 10^{-15}$.
These signal amplitude limits can be converted to limits on the luminosity distances or the number density of potential SMBBH systems.
For example, any binaries with chirp masses above $10^9M_{\odot}$ would have been detectable if they are located within a luminosity distance of 30 Mpc for gravitational wave frequency in the range from 10 to 200 nHz.
We constrain the number density of binaries with chirp masses around $10^8 M_{\odot}$ emitting gravitational waves at frequencies $\gtrsim \unit[10]{\rm nHz}$ to be no higher than 1 Mpc$^{-3}$.
We also constrain the SMBBH merger rate to be $R_{95}=4.4\times10^{-5} (10^{10} M_{\bigodot}/\mathcal{M})^{10/3}$ Mpc$^{-3}$Gyr$^{-1}$.

\begin{acknowledgments}

The Parkes radio telescope (Murriyang) is part of the Australia Telescope National Facility which is funded by the Australian Government for operation as a National Facility managed by CSIRO. We acknowledge the Wiradjuri People as the traditional owners of the Observatory site. 
This work was performed on the OzSTAR national facility at Swinburne University of Technology. The OzSTAR program receives funding in part from the Astronomy National Collaborative Research Infrastructure Strategy (NCRIS) allocation provided by the Australian Government, and from the Victorian Higher Education State Investment Fund (VHESIF) provided by the Victorian Government.
This paper includes archived data obtained through the Parkes Pulsar Data archive on the CSIRO Data Access Portal (http://data.csiro.au).
This work is supported by the National Natural Science Foundation of China (Grant No.~12203004), the National Key Research and Development Program of China (No. 2023YFC2206704), the Fundamental Research Funds for the Central Universities, and the Supplemental Funds for Major Scientific Research Projects of Beijing Normal University (Zhuhai) under Project ZHPT2025001.
Parts of this work were funded through the ARC Centre of Excellence for Gravitational Wave Discovery (CE230100016).
ZCC is supported by the National Natural Science Foundation of China under Grant No.~12405056, the Natural Science Foundation of Hunan Province under Grant No.~2025JJ40006, and the Innovative Research Group of Hunan Province under Grant No.~2024JJ1006. 

\end{acknowledgments}

\facilities{Parkes, OzSTAR}

\emph{Software:} \texttt{enterprise} \citep{enterprise}, \texttt{enterprise\_extensions} \citep{enterprise_extensions}, \texttt{PTMCMCSampler} \citep{justin_ellis_2017_1037579}, \texttt{libstempo} \citep{libstempo}, \texttt{tempo2} \citep{tempo2}, \texttt{astropy} \citep{astropy3, astropy2, astropy1}, \texttt{matplotlib} \citep{hunter2007matplotlib}, \texttt{healpy} \citep{zonca2019healpy}, \texttt{HEALPix} \citep{gorski2005healpix}.

\appendix
\label{sec:APPENDIX}

\section{Pulsar distances} \label{app:psr_distance}

In Table \ref{tab:distance}, we list the distance estimates for PPTA pulsars as used in the analysis while pulsar terms are taken into account.

\begin{table*}[t]
    \centering
    \begin{tabular}{ccccc|ccccc} 
        \toprule 
        Pulsar&Prior&Distance (kpc) & Error (kpc)$^\mathrm{*}$ & Ref. & Pulsar & Prior & Distance (kpc) & Error (kpc)$^\mathrm{*}$ & Ref.\\
        \midrule 
        J0030+0451 & PX   &0.329 & 0.005 & (1) & J0125–2327 & PX  & 1.2 & 0.2 & (2)\\
        J0437–4715 & PX   &0.156& 0.001 & (3) & J0613–0200 & PX  & 1.01 & 0.05 & (4) \\
        J0614–3329 & PX   &1.0 & 0.5 &  (5) & J0711–6830 & DM  & 0.11 & 0.02 &    \\
        J0900–3144 & DM   &0.38 & 0.08 &     & J1017–7156 & DM  & 1.8  & 0.4 &  \\
        J1022+1001 & PX   &0.72 & 0.02 & (6) & J1024–0719 & PX  & 1.08& 0.06 & (1)\\
        J1045–4509 & DM   &0.33 & 0.07 &  & J1125–6014 & PX  & 2 & 1 & (5) \\
        J1446–4701 & PX   &3 & 2 & (5)  & J1545–4550 & PX  & 5 & 3 & (5) \\
        J1600–3053 & PX   &1.39& 0.04  & (4) & J1603–7202 & DM  & 1.1 & 0.2 & \\
        J1643–1224 & PX   &0.94 & 0.09 & (1) & J1713+0747 & PX  & 1.07& 0.07 & (7) \\
        J1730–2304 & PX   &0.51 & 0.03 & (1) & J1744–1134 & PX  & 0.388& 0.005& (4) \\
        J1832–0836 & PX   &7 & 5 & (5)  & J1857+0943 & PX  & 1.15 & 0.08 & (4)\\
        J1902–5105 & DM   &1.7 & 0.3 &     & J1909–3744 & PX  & 1.158 & 0.003 & (8) \\
        J1933–6211 & PX   & 1.9 & 0.8 & (2)  & J1939+2134 & PX  & 3.0  & 0.3  & (1)\\
        J2124–3358 & PX   &0.48 & 0.02 & (4) & J2129–5721 & DM  & 6  & 1  &   \\
        J2145–0750 & PX   &0.63 & 0.02 & (6) & J2241–5236 & PX  & 1.05 & 0.05 & (9)\\
        \bottomrule 
    \end{tabular}
    \caption{The distances and their uncertainties for 30 PPTA pulsars. For pulsars labeled with a prior of PX, independent parallax measurements are available, with references: (1) \citet{vlbi_18psr_ding}, (2) \citet{psrdist_mpta}, (3) \citet{psrdist_2009_Deller}, (4) \citet{antoniadis2023second}, (5) \citet{psrdist_2025_ding}, (6) \citet{vlbi_57psr_deller}, (7) \citet{Chatterjee09}, (8) \citet{Liu20}, and (9) \citet{pptadr2_timing}. For pulsars without independent distance measurements, we used the dispersion measures (DM) provided in \citet{zic2023parkes}, and calculated distances based on the YMW16 \citep{yao2017new} model, and assigned an uncertainty of 20\%. Note that the assumed distance uncertainty is somewhat arbitrary and probably too small; however, we do not expect significant impact from pulsar distance priors on our upper limits.\\ $^\mathrm{*}$ $2^{+5}_{-1}$\,kpc for PSR~J1125$-$6014, while any other parallax-based pulsar distance reports the average of upper uncertainty and lower uncertainty as the distance uncertainty.}
    \label{tab:distance}
\end{table*}

\section{Posterior distributions of CGW parameters} \label{app:posterior}

\begin{figure}[h]
    \centering
    \includegraphics[width=\linewidth]{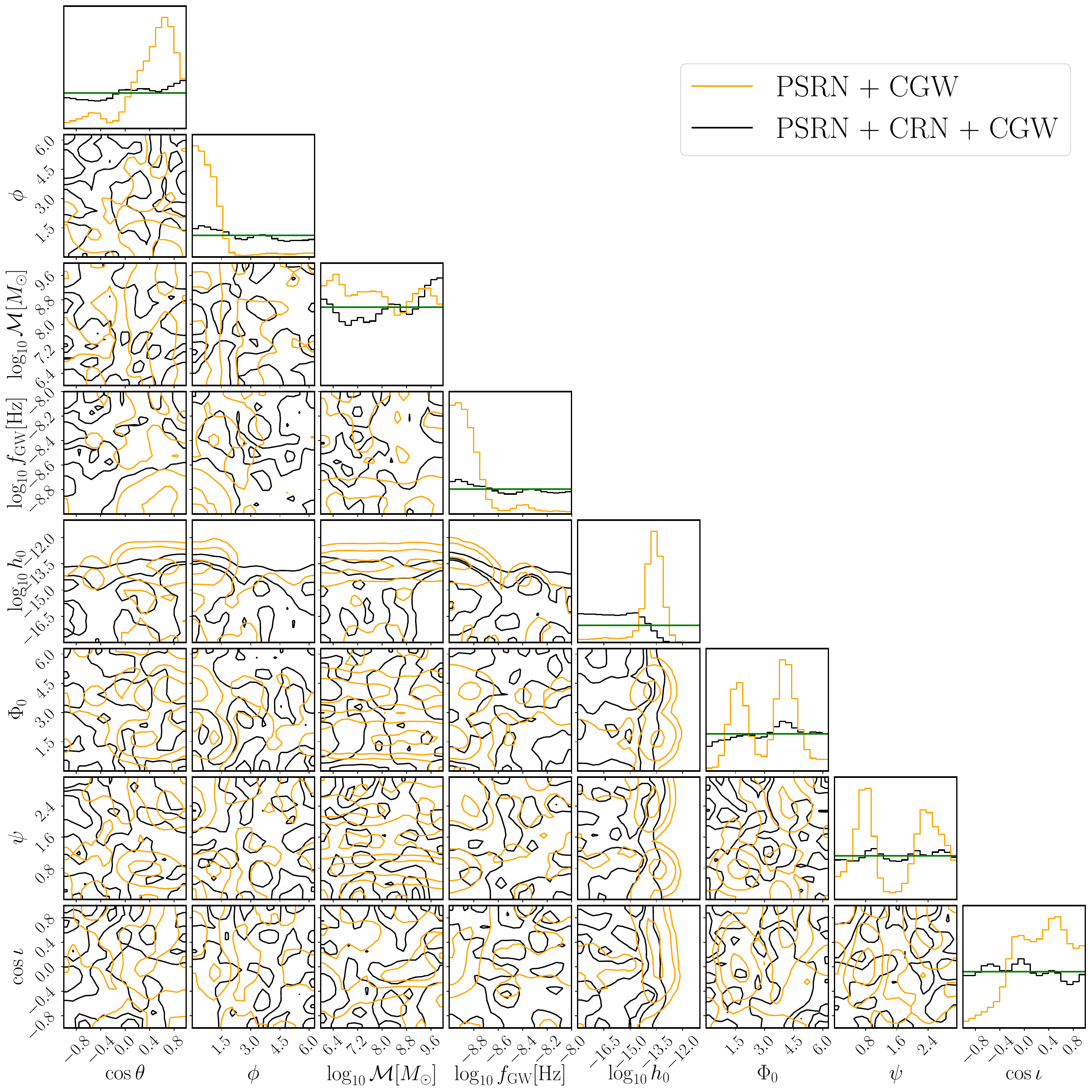}
    \caption{The posterior distributions of the CGW parameters show a clear peak at low frequency before taking the CRN into account (orange), which disappears after the inclusion of a CRN in the analysis (black). The green line is the prior distribution.}
    \label{fig:search_corner}
\end{figure}

Figure \ref{fig:search_corner} shows the posterior distribution of CGW parameters without (orange) or with (black) the inclusion of a CRN in the Bayesian search with PPTA DR3 data.
As one can see, pronounced peaks are present in the posterior distribution in the case of not accounting for a CRN, in agreement with our BFs shown in Figure \ref{fig:model_selection}. However, after accounting for a CRN, nearly all parameters, except for an upper limit on the signal amplitude, have posteriors as flat as their priors (plotted in green in the one-dimensional marginalized distribution).

One might be interested in the CGW posteriors under the HD-correlated GWB model instead of the CRN model. In this case, posterior distributions and Bayes factors can be calculated using the likelihood reweighting technique \citep{Payne19reweighting,reweighting}.
The weights used in this posterior reweighting procedure is given by the ratio of the HD-model likelihood to the CRN-model likelihood (assuming equal priors for both models),

\begin{equation}
w(d|\theta) = \frac{\mathcal{L}(d|\theta, \mathrm{HD})}{\mathcal{L}(d|\theta, \mathrm{CRN})}\, .
\label{eq:reweighting}
\end{equation}
The PPTA DR3 data set shows only weak evidence for HD correlations, 
with a Bayes factor of $\mathcal{B}^\mathrm{HD}_\mathrm{CRN} \approx 1.5$ over the CRN model \citep{reardon2023search}. 
Given the small Bayes factor, the posterior weights are close to unity. We therefore expect that adopting the HD-correlated GWB model would have a negligible effect on the CGW results presented in this work.

\section{The impact of CGW on the recovery of a CRN} \label{app:CW_ON_CRN}

\begin{figure}[h]
    \centering
    \includegraphics[width=0.5 \linewidth]{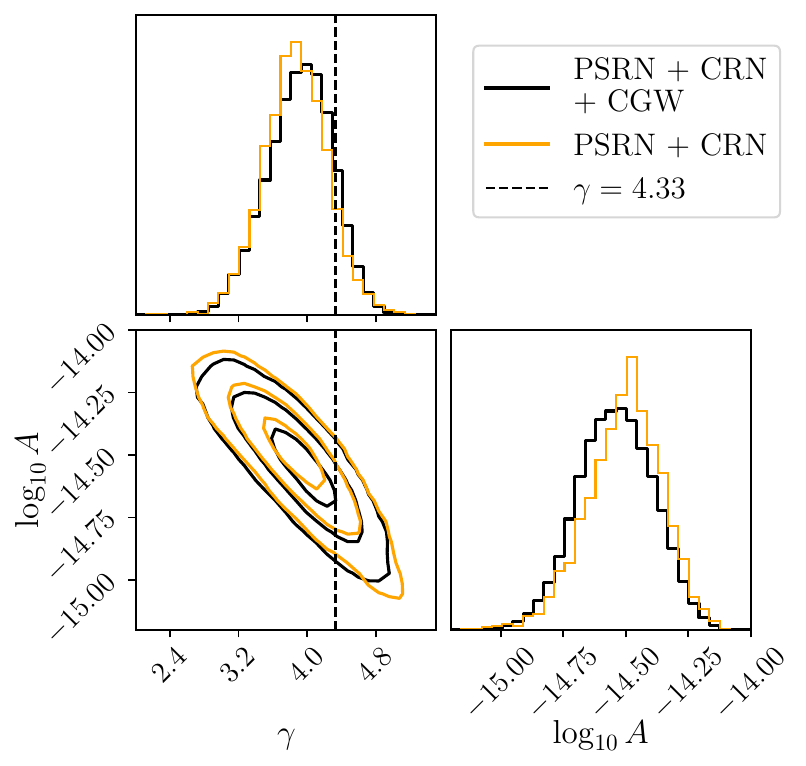}
    \caption{Posterior distributions of a power-law CRN parameters recovered using PPTA DR3 with (black) or without (orange) a CGW signal included in the analysis.}
    \label{fig:crn_comp}
\end{figure}

Figure \ref{fig:crn_comp} illustrates the impact of CGW on the estimation of a power-law CRN. It can be seen that the two sets of posteriors are nearly identical, indicating negligible support for the presence of a CGW. This is consistent with a small BF of 0.89 for PSRN+CRN+CGW against PSRN+CRN. 
The slight shift in the posterior distribution can be attributed to the covariance between a CGW and a CRN.
After including a CGW signal in the model, the best-fit $\gamma$ changes from 3.88 to 4.01, and $\log_{10} A$ changes from $-14.47$ to $-14.52$. This means that the overall strength of CRN has decreased due to the absorption of CRN power into a CGW.

\clearpage
\bibliographystyle{aasjournal}
\bibliography{refs}{}
\end{document}